%% file: main.tex
\documentclass[conference]{IEEEtran}
\IEEEoverridecommandlockouts
\usepackage{cite}
\usepackage{amsmath,amssymb,amsfonts}
\usepackage{algorithmic}
\usepackage{graphicx}
\usepackage{textcomp}
\usepackage{xcolor}

\usepackage{enumitem}
\usepackage{setspace}
\usepackage[margin=0.75in]{geometry}

\usepackage[labelformat=simple]{subcaption}

\newcommand{\titlename}{PhyCode}
\def\BibTeX{{\rm B\kern-.05em{\sc i\kern-.025em b}\kern-.08em
    T\kern-.1667em\lower.7ex\hbox{E}\kern-.125emX}}
\begin{document}

\title{\titlename: A Practical Wireless Communication System Exploiting Superimposed Signals}

\author{\IEEEauthorblockN{Wen Cui,
Chen Liu, Lin Cai and
Jianping Pan}
\IEEEauthorblockA{University of Victoria, BC, Canada}}
\vspace{-0.6cm}

\maketitle

\input{Abstract}
\input{Introduction}
\input{Relatedwork}
\input{SignalCollision}
\input{RecevierDesign}
\input{Performance}
\input{Conclusion}

\section*{Acknowledgement}
This work was supported in part by the Natural Sciences and Engineering Research Council of Canada (NSERC), Canada Foundation for Innovation (CFI), and BC Knowledge Development Fund (BCKDF).

\bibliographystyle{IEEEtran}  
\bibliography{main}

\end{document}

%% file: Abstract.tex
\begin{abstract}
	
Superimposed signals are anticipated to improve wireless spectrum efficiency to support the ever-growing IoT applications. 
Implementing the superimposed signal demands on ideally aligned signals in both the time and frequency domains. Prior work applied an average carrier-frequency offset compensation to the superimposed signal under the assumptions of homogeneous devices and static environments. However, this will cause a significant signal distortion in practice when heterogeneous IoT devices are involved in a dynamic environment. This paper presents \titlename, which exploits the nature of varying offsets across devices, and designs a dynamic decoding scheme which can react to the exact offsets from different signal sources simultaneously. 
We implement \titlename\ via a software-defined radio platform and demonstrate that \titlename\ achieves a lower raw BER compared with the existing state-of-the-art method.

\end{abstract}

%% file: Introduction.tex
\section{Introduction}
Wireless spectrum shortage is escalating with the growth of IoT applications. To solve this problem, a promising enabling technology is to allow wireless transmissions to overlap in the time/frequency/spatial domains, and decode 
the superimposed signals\footnote{In this paper, the superimposed signal refers to a signal that is generated from multiple signal sources and mixed in the same channel purposely.} to significantly increase the throughput of the wireless networks~\cite{zhang2006hot,katti2007embracing,chen2017bipass,zhang2017bi,zhang2017design,zhang2018design}. 
The theory underlying the superimposed signal has been around for several decades~\cite{shannon1961two}. Recent years have seen compelling advances in moving the superimposed signal from theory to practice~\cite{katti2007embracing,lu2013implementation}. Multiple systems have been implemented aiming to approach the theoretical throughput upper bound of wireless networks~\cite{chen2017bipass,kong2015mzig}. 



There are two common approaches for decoding the superimposed signal. One is successive interference cancellation~\cite{saito2013non}, which typically requires power control to guarantee that one signal has a much higher power than the other. Another approach is the physical-layer network coding~\cite{lu2013implementation} which is more promising for IoT devices since it does not require power control. However, PNC requires the symbol-level time synchronization, carrier-frequency synchronization, and carrier-phase synchronization. To do so, the existing systems~\cite{lu2013implementation,you2017reliable} required transmitters sharing a central clock (e.g., by GPS) to align their signals at the receiver to achieve the symbol-level time synchronization. For carrier-frequency synchronization and carrier-phase synchronization, they assumed homogeneous devices and static environments, where an average carrier-frequency offset can be compensated to different signal sources. 
\begin{figure}[!t]
	\centering
	\includegraphics[width=1.8in]{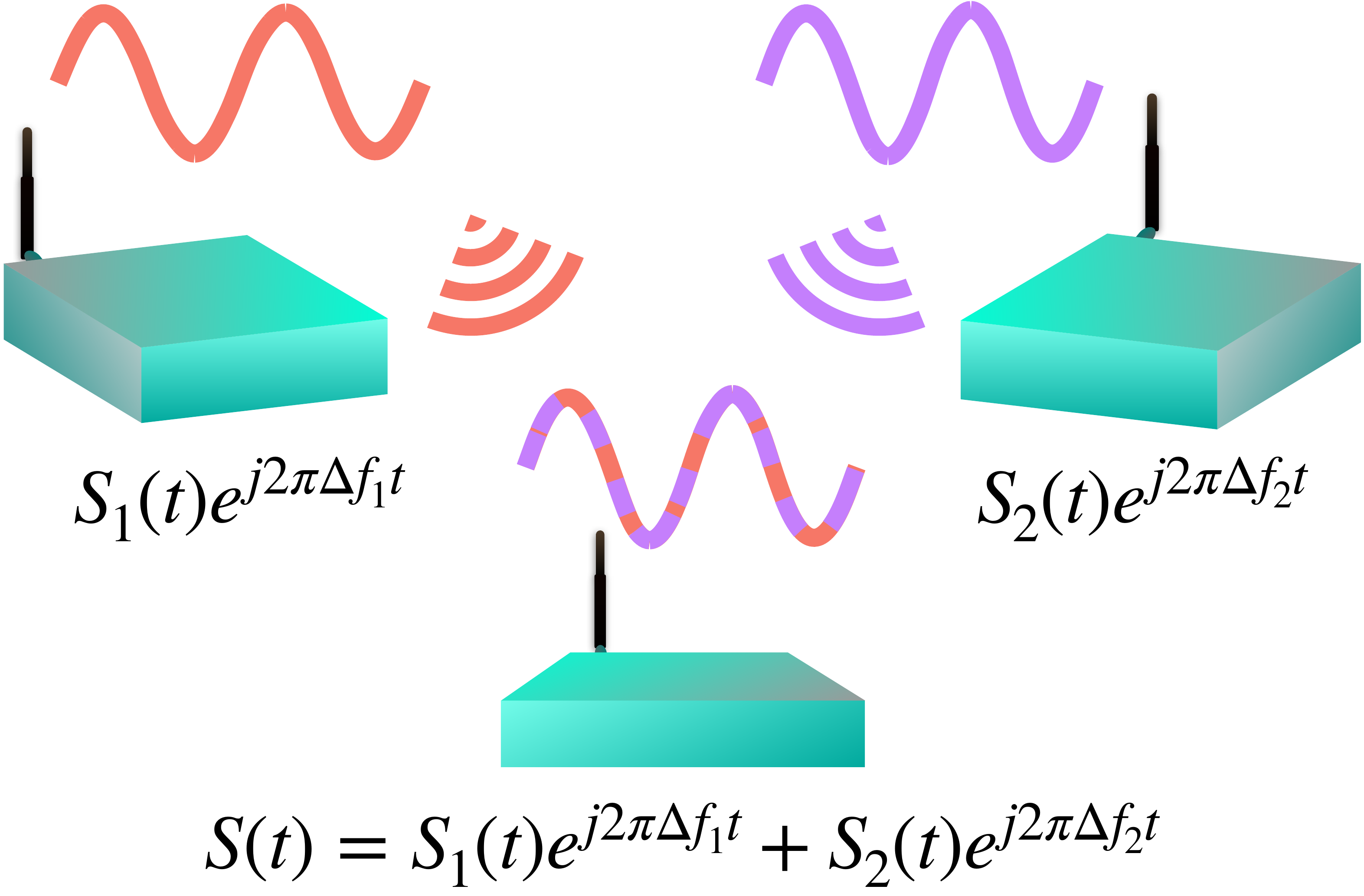}
	\captionsetup{font={footnotesize}}
	\caption{\textbf{A superimposed signal contains multiple offsets.} No single $\Delta f$ can be applied to $S(t)$ to compensate $\Delta f_{1}$ and $\Delta f_{2}$ simultaneously.}
	\label{fig_USRP}
	\vspace{-0.6cm}
\end{figure}
However, such an assumption may not hold in IoT scenarios, where heterogeneous devices operating in highly dynamic environments.
First, the IoT devices may vary in terms of vendors and types in a practical system. Specifically, the oscillator incorporated in the device inherently brings in a carrier frequency offset (CFO) when it converts signals from the carrier to the baseband~\cite{meyr1997digital}. Since oscillators are different from each other in terms of the crystal vibration characteristic\footnote{The different transmitters may have up to hundreds of kHz offset in the frequency domain caused by the oscillators. Even oscillators from the same producer can be different~\cite{lombardi2002fundamentals}.}, applying one compensation to two different CFOs simultaneously (as shown in Fig.~\ref{fig_USRP}) will lead to a high error rate. Additionally, different oscillators will cause a sampling frequency offset (SFO) when the receiver samples a signal, resulting in a remarkable phase shift to the signal. Therefore, the existing solutions are vulnerable in the real applications with heterogeneous devices. Second, IoT systems are typically deployed in dynamic environments with moving objects around (e.g., moving human, animal, or vehicle), and will naturally span large geographic areas full of NLOS scenarios. Under this condition, the theoretical detection accuracy of the superimposed signal cannot be maintained well, causing a symbol timing offset (STO). Therefore, the existing solutions are lack of robustness in dynamic environments.


This paper presents \titlename, a new approach to address the above limitations, aiming to enable the superimposed signal detection and decoding in practical systems. First, \titlename\ employs a two-step CFO correction scheme, where the CFO is calibrated in a coarse-grained manner at the transmitters and then the residual offset can be further corrected through a dynamic decoding scheme. By doing so, \titlename\ can react to the exact offsets from different signal sources simultaneously. Second, unlike the existing works which ignored the difference between oscillators in the sampling process, \titlename\ compensates SFO caused by different oscillators, which makes the design ubiquitous to the device heterogeneity. Third, \titlename\ employs a two-step correlation method for the superimposed signal detection, by which the computation cost can be reduced significantly and the corresponding detection accuracy is improved under a practical NLOS scenario. Fourth, \titlename\ corrects STO that is caused by time synchronization error in a dynamic environment. Finally, \titlename\ exploits the nature of varying offsets, and designs a dynamic decoding scheme. Thus, \titlename\ is robust to dynamic environments.

To the best of our knowledge, this paper is the first design to implement dynamic offset tracking and reacting schemes to detect and decode superimposed signals, and thus to achieve a lower BER in practical systems. \titlename\ focuses on OFDM, implementable for IEEE 802.11 a/g/n/p systems
, LTE, etc. \titlename\ can be a key enabling technology for many promising wireless technologies requiring the decoding of superimposed signals, such as non-orthogonal multiple access (NOMA), and physical-layer network coding (PNC). 
We summarize our contributions as follows:

\begin{itemize}[noitemsep,leftmargin=*]
	\item We present a new approach of implementing superimposed signals that considers noises from synchronization, oscillator offsets, and dynamic environments. As the result, the design is robust to real deployment without the assumptions of homogeneous devices and static environments.
	\item We demonstrate our design on a software-defined radio platform. We further design our system to be compatible to IEEE 802.11-like systems with only minor changes.
	\item We conduct extensive experiments with our prototype. The results revealed that our design can effectively mitigate the impact of the offsets and achieve a much lower raw BER, compared to the existing state-of-the-art method.
\end{itemize}

%% file: Relatedwork.tex
\section{Related Work}
Prior work falls into the following three categories.

{\bf (a) Successive Interference Cancellation in NOMA:} Interference cancellation schemes typically require power control to guarantee that one signal has a much higher power than the others. In this case, they can decode one first, and then cancel it out and decode the others. Such a design has been proposed for cellular systems~\cite{saito2013non}. Unlike interference cancellation schemes, \titlename\ is a new type of NOMA which does not need power control on interfered signals, as controlling power tightly is hard for lightweight/ubiquitous IoT devices in most of the cases~\cite{mahinthan2008partner}.

{\bf (b) Physical-layer Network Coding (PNC):} PNC~\cite{lu2013implementation} can operate on superimposed signals without power control. \titlename\ can also be used for the PNC implementation in IoT systems. Theoretically, PNC requires symbol-level time synchronization, carrier-frequency synchronization, and carrier-phase synchronization. The existing PNC implementation~\cite{you2015network} used a GPS clock to align the transmitted signals at the receiver to achieve the symbol-level synchronization in the time domain (more details are discussed in Section~\ref{sec:t_syn}). Therefore, the main challenges are to develop solutions to address the carrier-frequency synchronization and carrier-phase synchronization issues. Although several approaches can achieve a tight synchronization in the frequency domain, the phase domain~\cite{hamed2018chorus} and the time domain~\cite{rahul2010sourcesync,vasisht2016decimeter}, they require either sophisticated hardware or a strict scheduler, which will introduce onerous burden for implementing PNC, especially in IoT systems. To solve this problem, the prior work~\cite{lu2013implementation,you2017reliable,you2015network} compensated an average carrier-frequency offset to different signal sources under the assumption of homogeneous devices and static environments. However, 
such an assumption may not hold, which limits the superimposed signal detection and decoding in ubiquitous and robust IoT systems. Different from the existing solutions, \titlename\ takes the heterogeneous devices and dynamic environments into consideration; therefore, \titlename\ can dynamically react to the exact offsets from different signal sources simultaneously. 

Some other related work falls in the area of network coding in the physical layer. One is analog network coding~\cite{katti2007embracing}. Although this solution does not require time synchronization, the relay node not only amplifies the received signal, but also amplifies the noise, which causes error propagation. Another is BiPass~\cite{chen2017bipass}, which has also been investigated to improve the throughput by decoding the superimposed signal; however, it requires dedicated full-duplex devices, and more importantly, it still suffers from the noise propagation issue.


{\bf (c) Superimposed Signal in Other Techniques:} Decoding superimposed signal has also been widely adopted in other popular wireless techniques, such as RFID~\cite{wang2012efficient, jin2017fliptracer, jin2018parallel}, Zigbee~\cite{ kong2015mzig} and LoRa~\cite{eletreby2017empowering,hessar2018netscatter}. However, these techniques were designed to transmit at a low data rate, which limits wireless spectrum efficiency. In contrast to these works, \titlename\ focuses on OFDM, implementable for IEEE 802.11 a/g/n/p, LTE, etc., which enables higher spectrum efficiency. 

%% file: SignalCollision.tex
\section{Preliminary}
\label{sec:t_syn}

\begin{figure}[!t]
    \centering
    \vspace{0.035in}
    \includegraphics[width=2.1in]{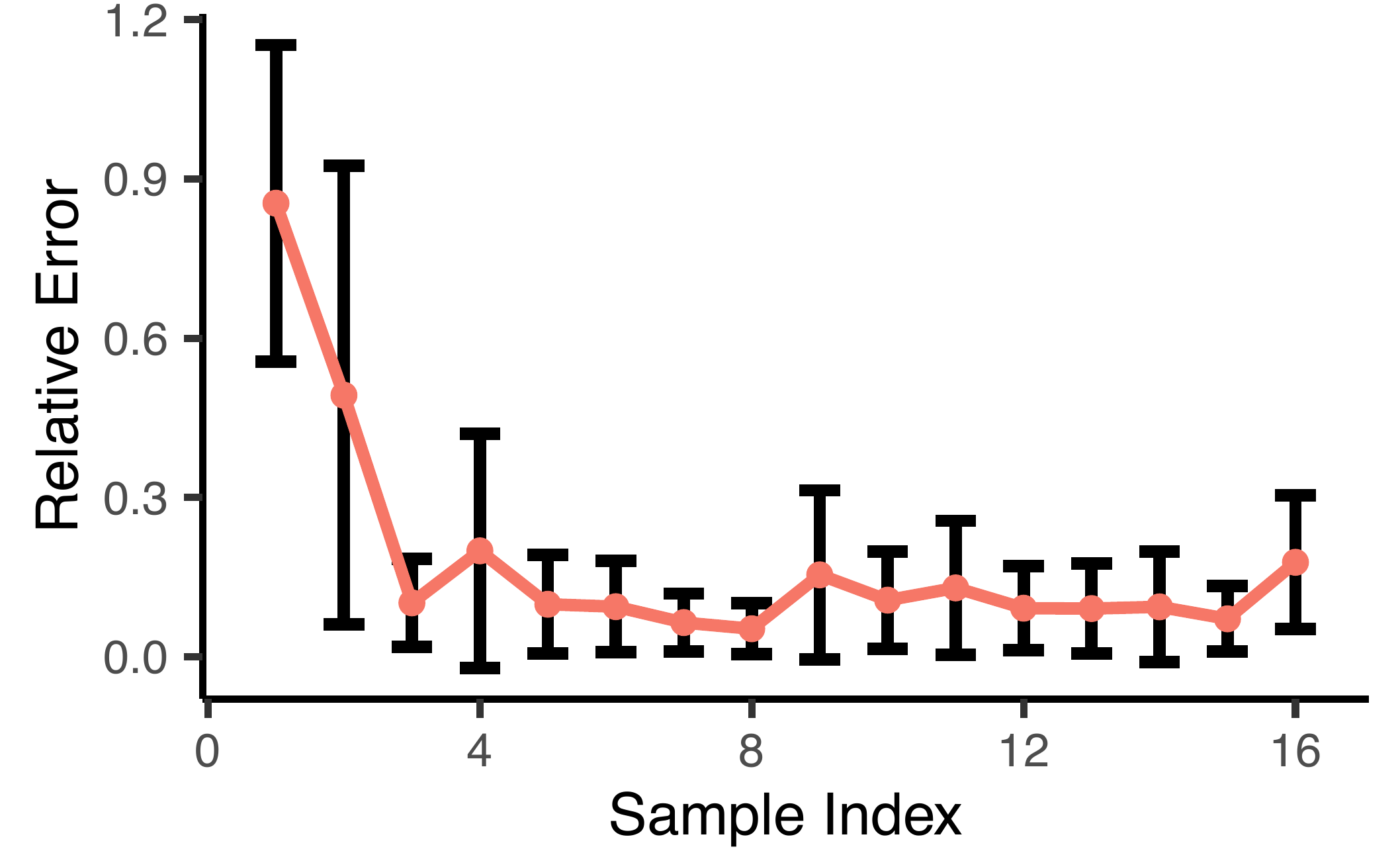}
    \captionsetup{font={footnotesize}}
    \setlength{\abovecaptionskip}{-0.05cm}
    \caption{\textbf{Multipath effect on CP.} Delayed multipath signals appear only in the first two samples of the CP part.}
    \label{fig_cp_multipath}
    \vspace{-0.7cm}
\end{figure}

In this section, we present some background information and discuss the time synchronization requirement for decoding the superimposed signal. 
For simplicity, we consider two sources transmitting signals simultaneously. Let $Y$ be the received signal, $X_{1}$ and $X_{2}$ the transmitted signals, and $H_{1}$ and $H_{2}$ the corresponding channels between the two transmitters and receiver, respectively. For notation simplicity, we represent the received superimposed signal as

\begin{equation}
Y = H_{1}X_{1} + H_{2}X_{2},\label{equ:re}
\end{equation}
in the frequency domain. Note that the above representation is only valid in an ideal scenario. When it comes to practice, the time synchronization, oscillator offsets, and environment noises should be taken into account. We will address these issues in this section and Section~\ref{sec:design}.

Two questions should be concerned before we design a practical system for decoding superimposed signals: (\textrm{i}) what is the required time synchronization level? (\textrm{ii}) can we achieve that level of synchronization accuracy on off-the-shelf devices?


We can address the first question by considering the usable number of cyclic prefix (CP) samples. A superimposed signal experiences the same multi-path fading channel as a single source signal does, since each signal has its own multipath components. Thus, not all the samples in CP can be used to adjust the fast Fourier transform (FFT) window due to inter-symbol interference (ISI). To see it clearly, we conducted a benchmark experiment in an office with rich multipath. As shown in Fig.~\ref{fig_cp_multipath}, most of the multipath signals arrive at the receiver within the first 2 sample intervals, which is consistent with the previous observations~\cite{tse2005fundamentals}. In this case, the rest 14 sample intervals can be utilized to adjust the FFT window. Specifically, for IEEE 802.11a and 802.11p, the time duration of CP is 0.8$\mu s$ and 1.6$\mu s$, respectively. Accordingly, the time duration of 14 samples is 700$ns$ and 1400$ns$, respectively. Hence, the required time synchronization for the implementation of the superimposed signal is about hundreds of $ns$.


\begin{figure}[!t]
	\centering
	\vspace{0.05in}
	\includegraphics[width=2.5in]{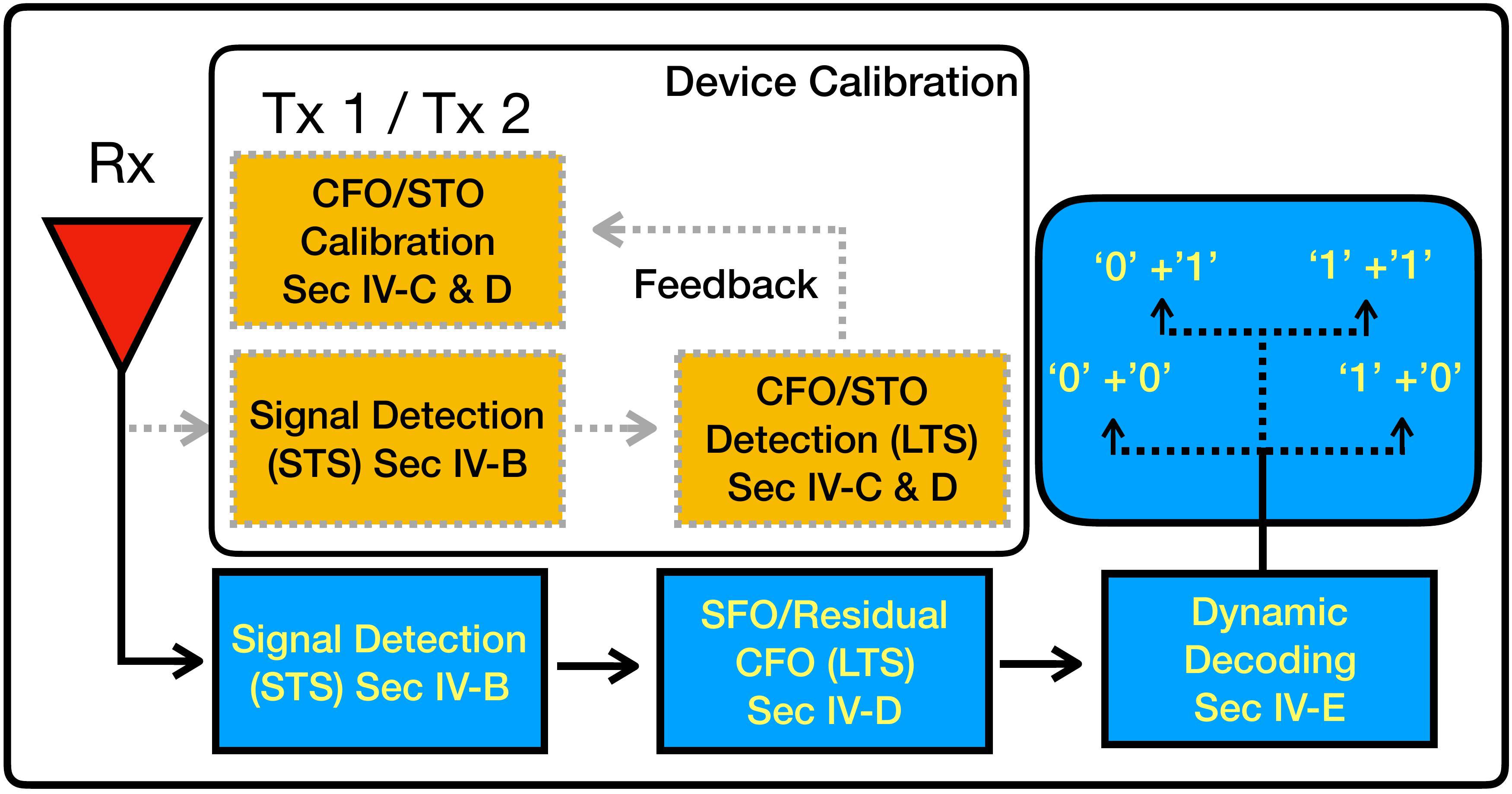}
	\captionsetup{font={footnotesize}}
	\caption{\textbf{\titlename\ overview.}}
	\label{fig_overview}
	\vspace{-0.6cm}
\end{figure}

For the second question, many synchronization techniques have been developed recently, making this level of synchronization achievable on off-the-shelf devices. For instance, devices can maintain around 300$ns$ accuracy by only using a GPS clock~\cite{schmitz2017distributed}. Thus, current time synchronization techniques provide a solid foundation for implementing the superimposed signal on off-the-shelf devices. In this paper, we focus on the unsolved problems that affect the practical implementation of the superimposed signal decoding, i.e., heterogeneous devices and dynamic environments with NLOS scenarios.

%% file: RecevierDesign.tex
\section{Design of \titlename}
\label{sec:design}


In this section, we introduce the design of \titlename. Four main modules of \titlename\ have been shown in Fig.~\ref{fig_overview}. 
\subsection{Preamble Design}\label{sec:preamble}
In a Wi-Fi system, every packet starts with a preamble including a short training sequence (STS) and a long training sequence (LTS). STS is used for signal detection, while LTS is designed for measuring the difference between the received and transmitted pilot. For the superimposed signal, STS can still be used for signal detection (see Section~\ref{sec:detection}). However, collision makes the existing solutions infeasible to measure those differences based on LTS. Naively, we can design an orthogonal LTS on signals from each source. 
In this case, when two signals arrive at the receiver, the differences can be measured separately. Unfortunately, it is infeasible to guarantee perfect orthogonality due to the device and channel variations in practice.
	
\begin{figure}[!t]
	\centering
	\vspace{0.05in}
	\includegraphics[width=2.3in]{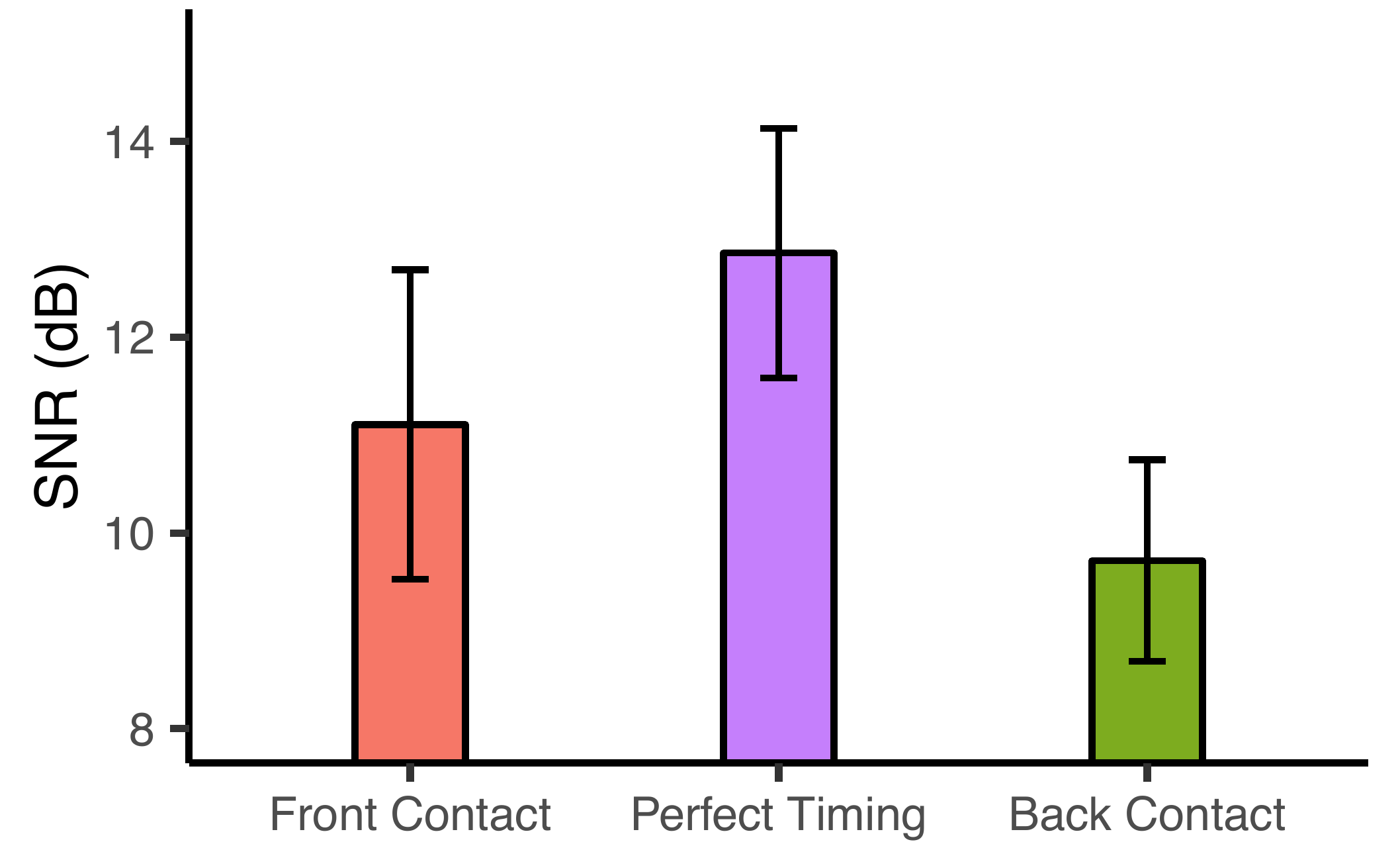}
	\captionsetup{font={footnotesize}}
	\setlength{\abovecaptionskip}{-0.05cm}
	\caption{\textbf{SNR under different positions of the FFT window.}}
	\label{fig_moving_fft_window}
	\vspace{-0.3cm}
\end{figure}

\begin{figure}[!t]
	\centering
	\includegraphics[width=2.3in]{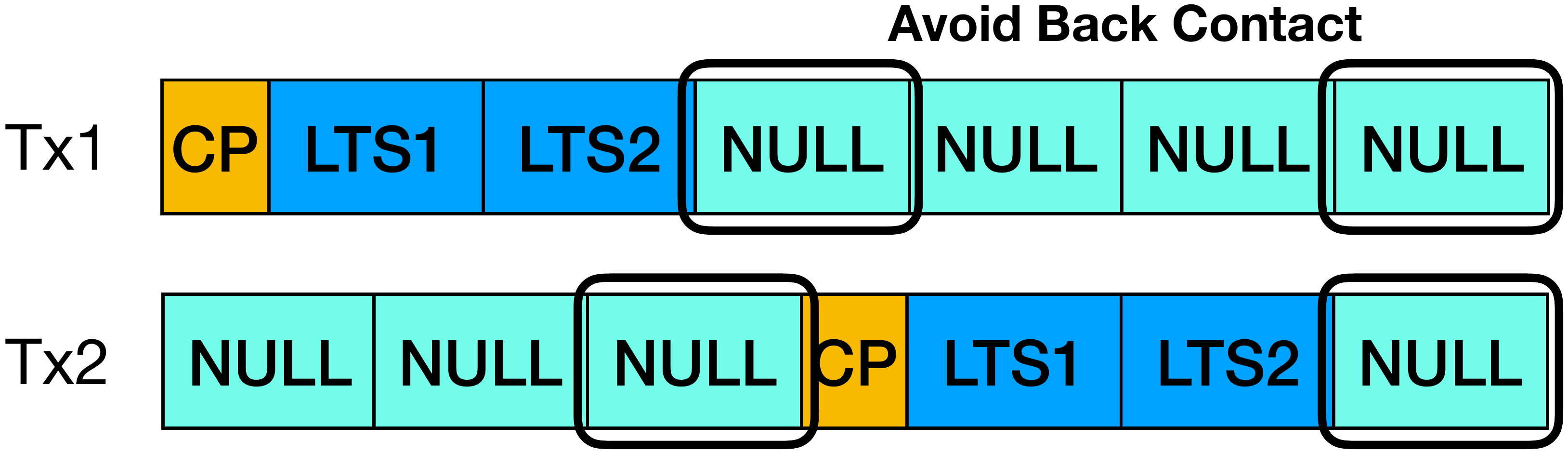}
	\captionsetup{font={footnotesize}}
	\caption{\textbf{LTS design for the superimposed signal.}}
	\label{fig_fft_window_double}
	\vspace{-0.5cm}
\end{figure}

To analyze the effect of this non-perfect orthogonality on decoding performance, such as SNR, we further conduct an experiment where we artificially introduce latency for different sources. For simplicity, we define LTS collision as ``contact''. In particular, {\em back contact} refers to the case where samples from other sources are included in the FFT process, while {\em front contact} means that a sliding FFT window only contains its own CP and data samples. The experiment result is shown in Fig.~\ref{fig_moving_fft_window}. It is observed that the perfect FFT position gives us the best SNR, while the {\em front contact} induces a phase shift to the signal. In contrast, the {\em back contact} gives us the worst SNR. Note that the phase shift caused by the {\em front contact} can be corrected later through STO calibration (see Section.~\ref{sec:timeoffset}). However, the SNR degradation induced by the {\em back contact} cannot be corrected in further steps. This observation implies that it is needed to avoid the {\em back contact} cases. To do this, we design our own LTS as described in Fig.~\ref{fig_fft_window_double}. Different from directly applying orthogonal LTS, we insert two extra NULL symbols to enlarge the distance of two sources in the time domain, which can successfully avoid the {\em back contact} in most of the practical cases.

\subsection{Superimposed Signal Detection}\label{sec:detection}

\begin{figure}[!t]
	\begin{minipage}[t]{0.5\linewidth}
		\centering
		\includegraphics[width=4.2cm]{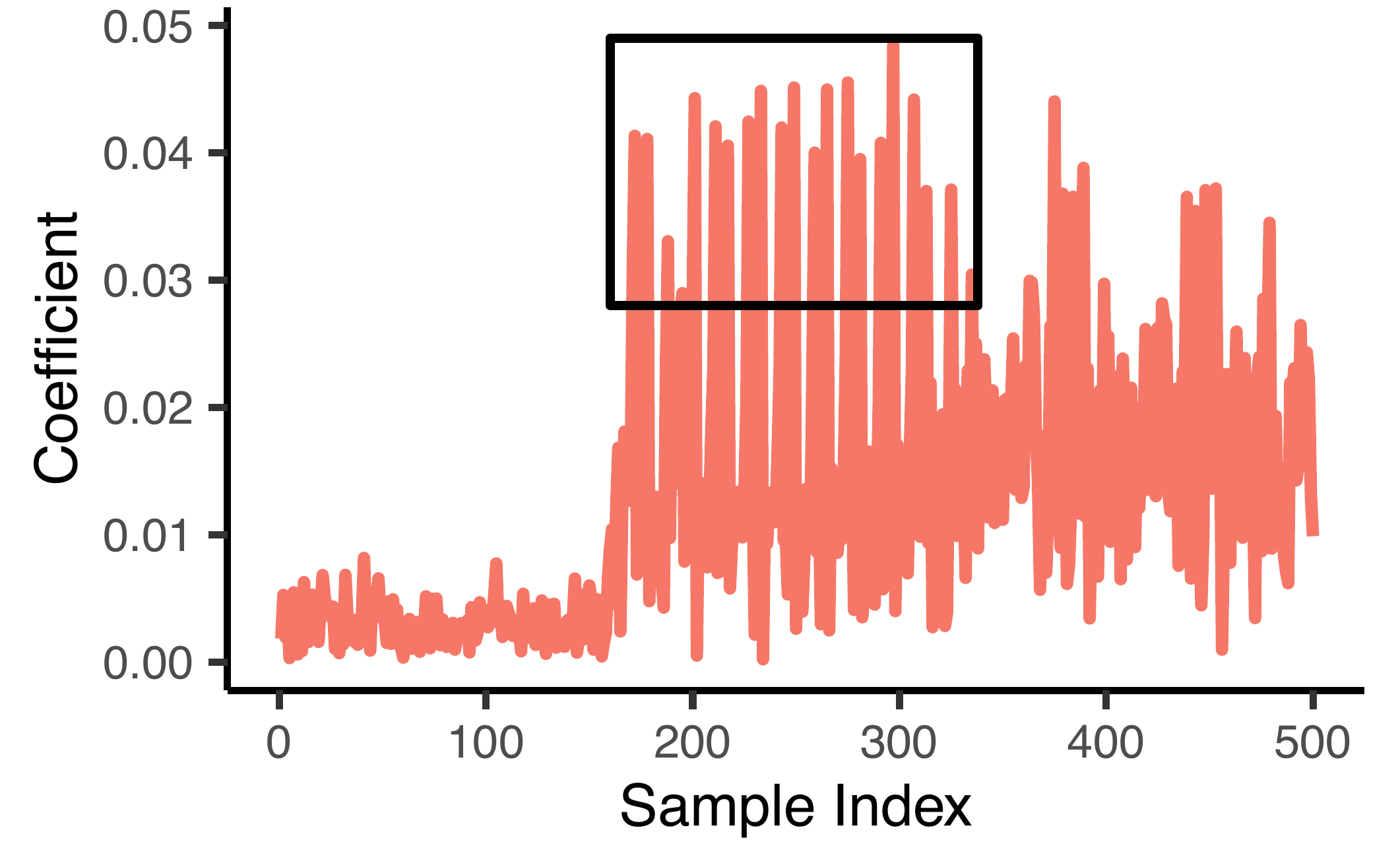}
		\captionsetup{font={footnotesize}}
		\caption{\textbf{Cross-Correlation.}}
		\label{fig_STS_unsync_cross}
	\end{minipage}%
	\hspace{-0.5cm}
	\begin{minipage}[t]{0.5\linewidth}
		\centering
		\includegraphics[width=4.2cm]{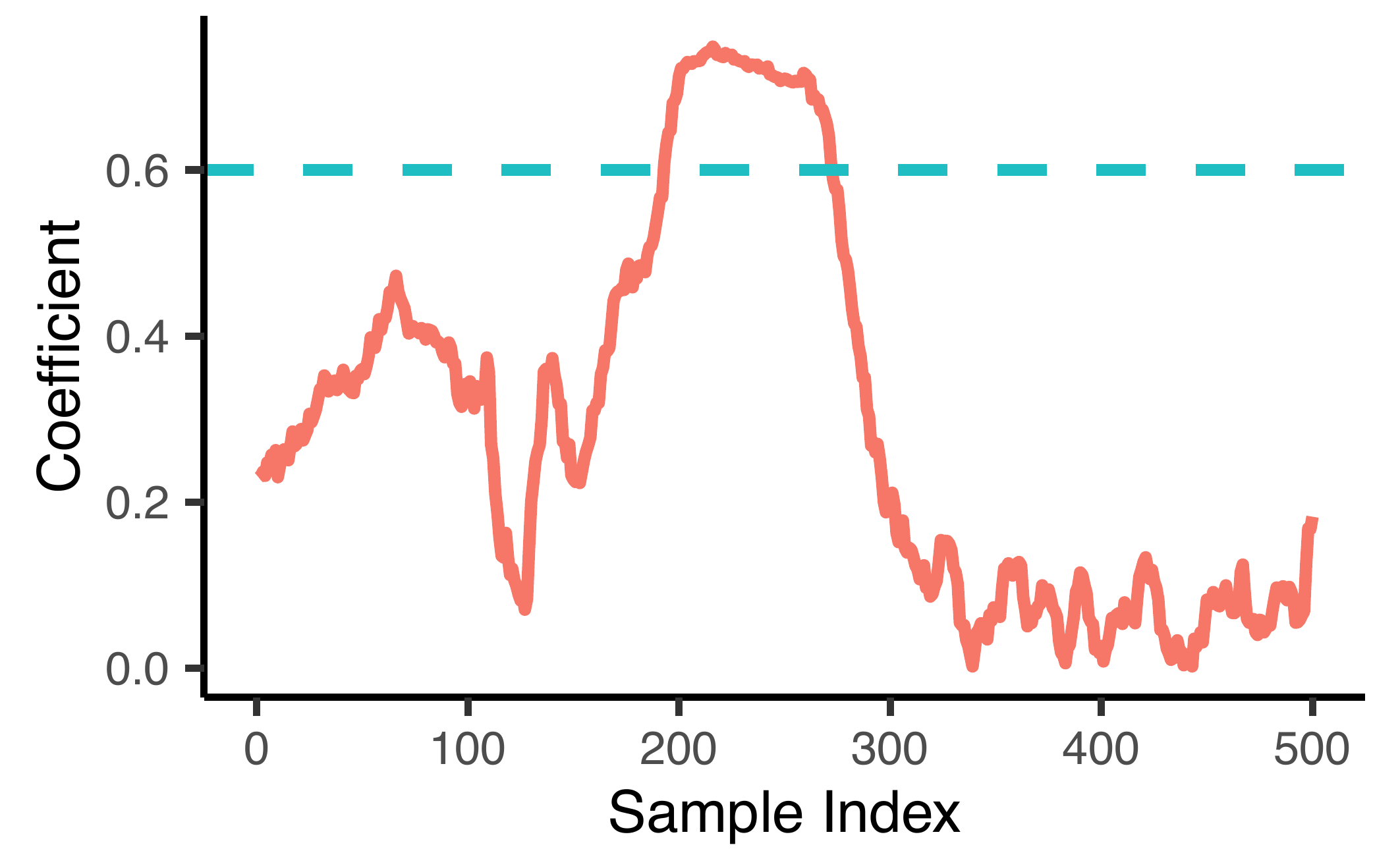}
    	\captionsetup{font={footnotesize}}
		\caption{\textbf{Auto-Correlation.}}
		\label{fig_STS_unsync_auto}
	\end{minipage}%
	\vspace{-0.4cm}
\end{figure}

In dealing with the superimposed signal detection, cross-correlation has been widely adopted to obtain a high detection accuracy~\cite{GollakotaZigzag}. The cross-correlation involves every coming sample 
to execute complex multiplication with a significant computation cost (e.g., 64 times of complex multiplication for each sample in 802.11a/p). {\em However, when it comes to practice, there are two challenges remaining.} First, this theoretical detection accuracy cannot maintain anymore in the practical dynamic environment where the channel is complicated. For example, Fig.~\ref{fig_STS_unsync_cross} illustrates the correlation pattern of a superimposed signal in a common indoor environment with rich multipath. As we can see, the correlation pattern is not so clear to easily distinguish the two signals, leading to a low detection accuracy. Second, a high computation cost would lead to a low performance of the whole system~\cite{bloessl2018performance}, which will eventually hurt the decoding accuracy (i.e., incur many more error bits) at the receiver.

\begin{figure*}
	\begin{minipage}[t]{0.33\linewidth}
		\centering
		\includegraphics[width=5.3cm]{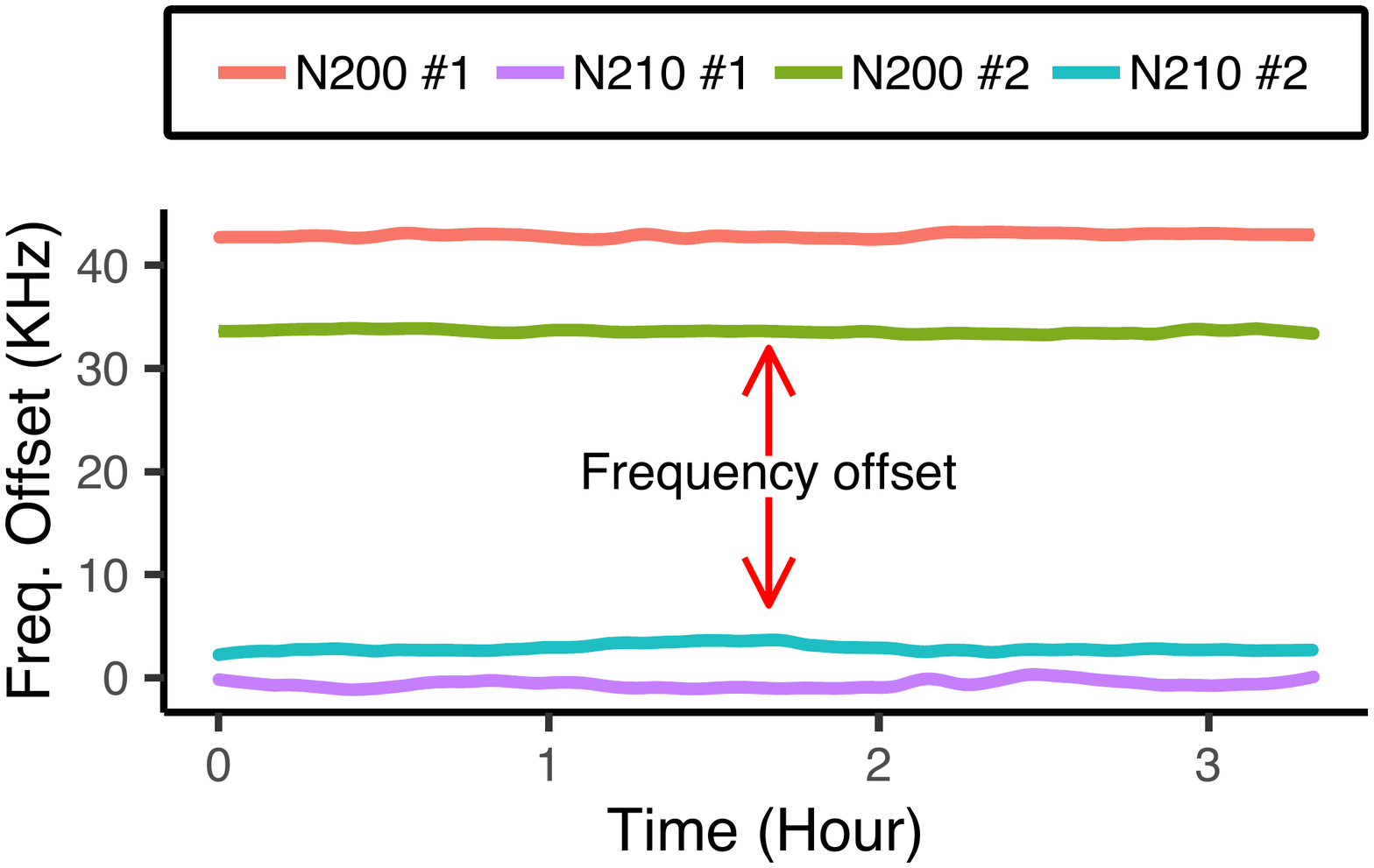}
			\captionsetup{font={footnotesize}}
		\caption{\textbf{Carrier frequency offset.}}
		\label{fig_CFO}
	\end{minipage}%
    \begin{minipage}[t]{0.33\linewidth}
	\centering
	\includegraphics[width=5.4cm]{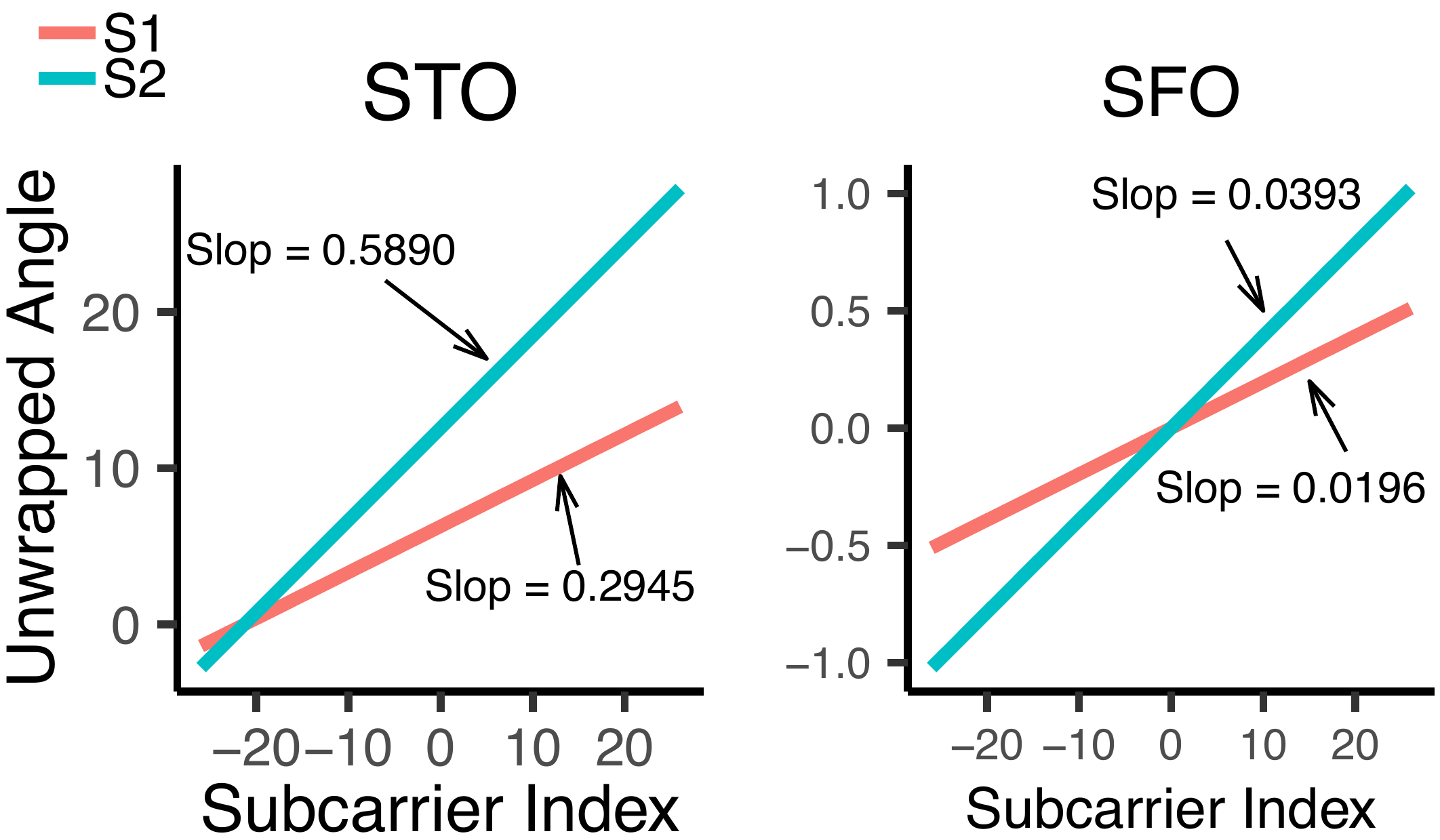}
	\captionsetup{font={footnotesize}}
	\caption{\textbf{Understanding of STO and SFO.}}
	\label{fig_STO_SFO}
    \end{minipage}%
    \begin{minipage}[t]{0.33\linewidth}
	\centering
	\includegraphics[width=5.3cm]{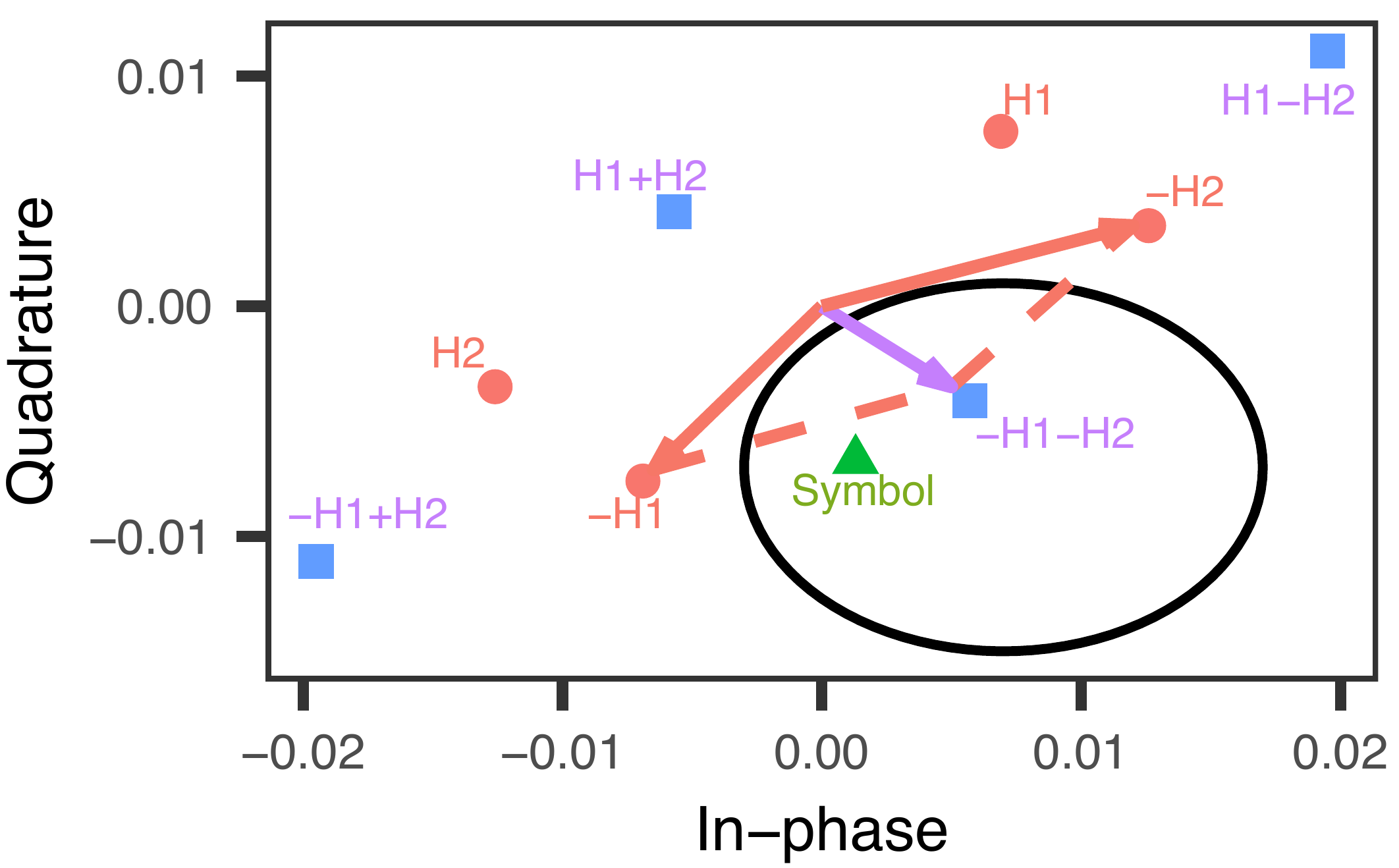}
	\captionsetup{font={footnotesize}}
	\caption{\textbf{Constellation map.}}
	\label{fig_constellation}
    \end{minipage}%
	\vspace{-0.4cm}
\end{figure*}

In order to solve these two challenges, we propose a two-step correlation method with a low computation cost and a comparable detection accuracy. In the first step, we apply auto-correlation~\cite{schmidl1997robust} with STS to detect the signals in a coarse-grained manner (e.g., only one complex multiplication for each sample). By doing so, the computation cost can be reduced significantly. In the following step, we design a cross-correlation only for LTS to further improve the detection accuracy under the complicated channel. We plot the result of a benchmark experiment in Fig.~\ref{fig_STS_unsync_auto}. Clearly, this two-step correlation can achieve a comparable detection accuracy as the cross-correlation 
in a real-world scenario.


\subsection{Carrier Frequency Offset Calibration}%
To understand the side effect of synchronization error, we start from a single source signal. Typically, losing synchronization will distort the signal from three main respects: STO denoted by $n_{\epsilon}$, CFO denoted by $\Delta f$, and SFO denoted by $\Delta T$. Specifically, the distorted signal $r(t_{n})$ can be represented as below,
\begin{equation}
r(t_{n}) = e^{j2\pi \Delta f n T} \sum_{i}h_{i}(nT')(s(n-n_{\epsilon})T'-\tau_{i}) + n_{0},
\label{eq_rx}
\end{equation}
where $T'$ and $T$ are the sampling time at the receiver and the transmitter, respectively~\cite{meyr1997digital} (the difference between $T'$ and $T$ will cause SFO). We denote $h_{i}$ as the channel impulse response, $\tau_{i}$ as the delay, and $n_{0}$ refers to the white Gaussian noise. In the superimposed signal, however, the synchronization error becomes even worse as multiple offsets from different sources are involved. Here, we focus on addressing CFO first, and then STO and SFO will be analyzed in the next subsection.


As mentioned early (see Fig.~\ref{fig_USRP}), suppose that there are two transmitted signals with CFO $\Delta f_{1}$ and $\Delta f_{2}$, respectively. Therefore, applying one compensation to two different CFOs simultaneously will lead to a high error rate. 
To design a practical implementable system, we must deal with the device heterogeneity and compensate $\Delta f_{1}$ and $\Delta f_{2}$, respectively. To do this, we design a two-step CFO correction scheme. In the first step, we calibrate the CFO in a coarse-grained manner at the transmitters, while during the second step, the residual offset can be further corrected by applying the dynamic decoding scheme. In this case, the CFO can be divided into two parts: $\Delta f_1=\Delta f_{1_{\rm{ step1}}} + \Delta f_{1_{\rm{step2}}}$ ($\Delta f_2$ can also be written like this). We first focus on the first step and the details of the second step are discussed in Section~\ref{sec:dyn}.

To design the first step of CFO calibration, we conduct a three-hour experiment in an ordinary office with different devices, including one USRP N210 as the transmitter and four USRPs as the receivers (two N210s and two N200s). As we can notice from Fig.~\ref{fig_CFO}, a large amount of CFO exists in different devices, especially when their types are different. However, it is very interesting to observe that the CFO differences are considerably stable even after a few hours. Hence, we leverage this observation, and compensate this CFO (i.e., $\Delta f_{1_{\rm{step1}}}$) beforehand in a coarse-grained manner. After this calibration, only small CFO still left, and based on that, we further develop the second step for fine-grained correction.

\subsection{Timing Offset Calibration}
\label{sec:timeoffset}

Besides CFO, the remaining offsets are STO and SFO.  
Intuitively, STO and SFO are both timing problems. 
The difference is that STO comes from the receiving process, such as limited computation power, noisy circuit, etc., which could introduce a few samples latency. On the other hand, SFO comes from the oscillator, sharing the same reason with CFO. 
Although STO and SFO are caused by different reasons, they both induce a phase shift to the signal. Here, we define $\theta^{STO}$ and $\theta^{SFO}$ for the phase shifts caused by STO and SFO, respectively. The phase shifts in one symbol can be described as
\begin{equation}
\theta^{STO}_{k} = 2\pi k n_{\epsilon}/N_{c},
\label{eq_single_sto}
\end{equation}
\begin{equation}
\theta^{SFO}_{k} = 2\pi k \gamma (N_{c}+L)/N_{c},
\label{eq_single_sfo}
\end{equation}
where $\gamma=(T-T')/T$ is defined as the sampling time error ratio, and $k$ is the subcarrier index, $L$ the length of the CP, and $N_{c}$ the length of the data part in every symbol. 


\begin{figure*}
	\centering
	\begin{subfigure}{.2\textwidth}
		\centering
		\includegraphics[width=1.4in]{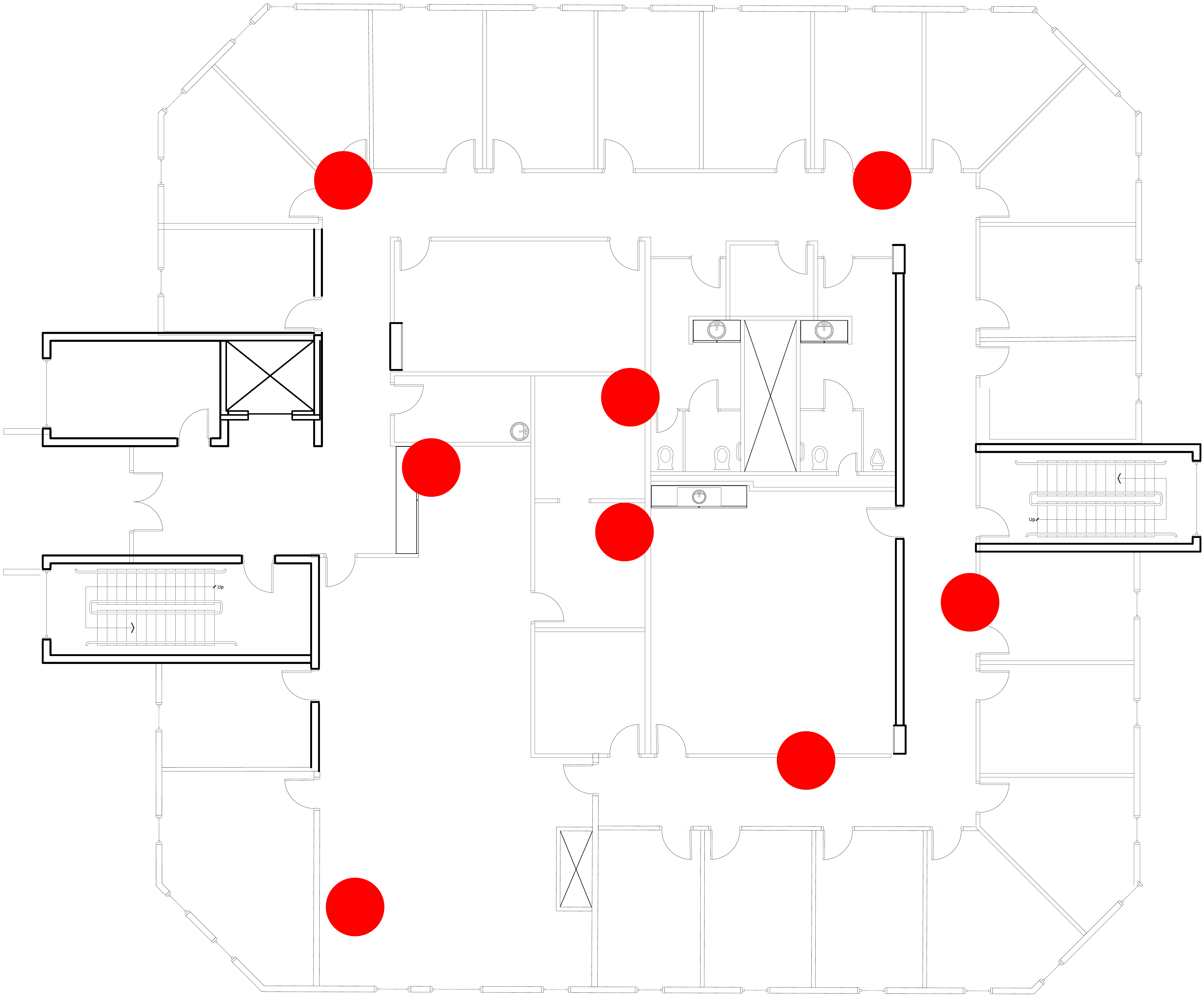}
		\captionsetup{font={scriptsize}}
		\caption{\textbf{The deployment layout}}
		\label{fig_exp_floor}
	\end{subfigure}
	\hspace{.2cm}
	\begin{subfigure}{.2\textwidth}
		\centering
		\includegraphics[width=1.43in]{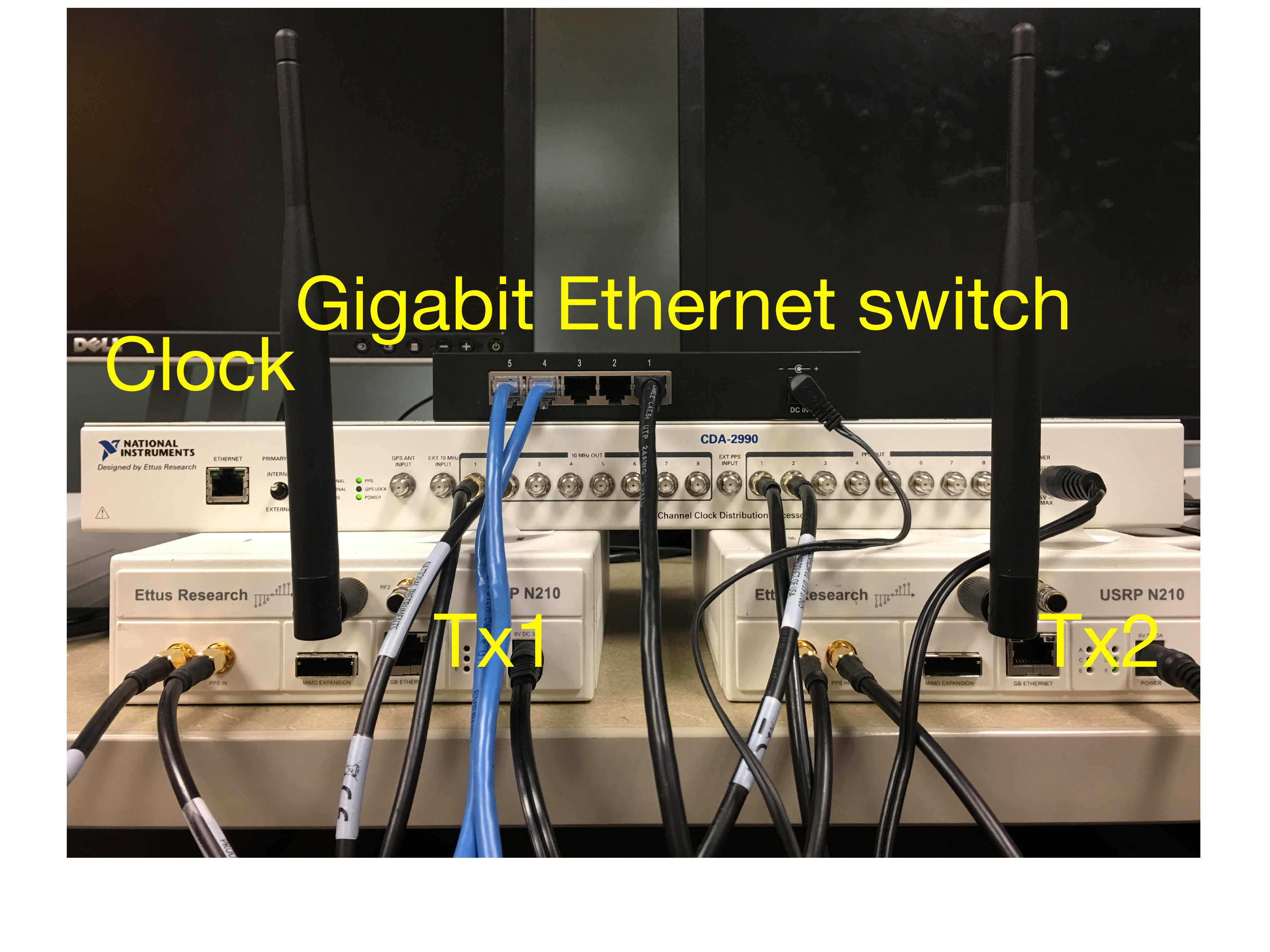}
		\captionsetup{font={scriptsize}}
		\caption{\textbf{Transmitters}}
		\label{fig_exp_tx}
	\end{subfigure}
	\hspace{.2cm}
	\begin{subfigure}{.2\textwidth}
		\centering
		\includegraphics[width=1.43in]{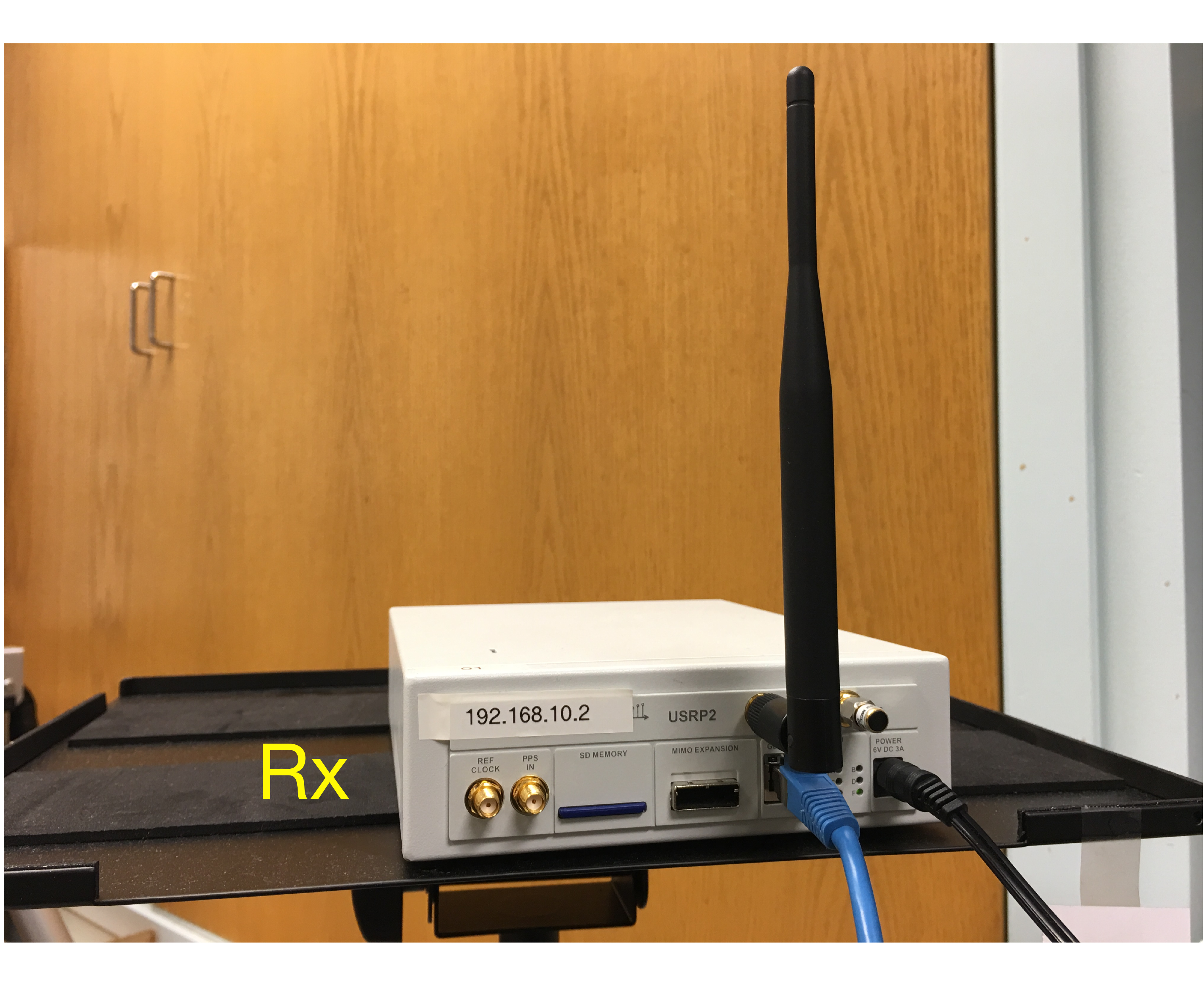}
		\captionsetup{font={scriptsize}}
		\caption{\textbf{Receiver}}
		\label{fig_exp_rx}
	\end{subfigure}
	\hspace{.0cm}
	\begin{subfigure}{.2\textwidth}
		\centering
		\includegraphics[width=1.7in]{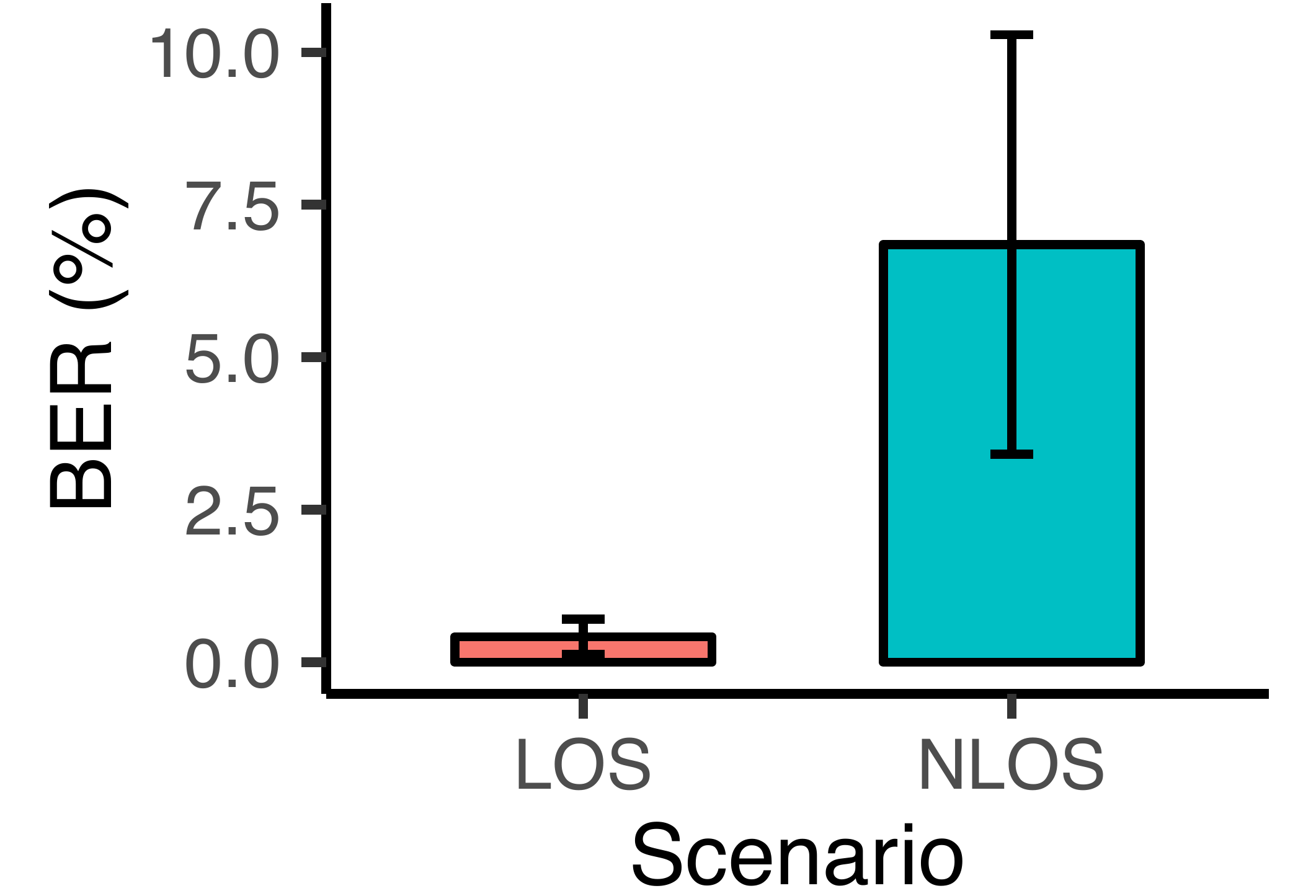}
		\captionsetup{font={scriptsize}}
		\caption{\textbf{BER in dynamic environments}}
		\label{fig_exp_scenario}
	\end{subfigure}
	\captionsetup{font={footnotesize}}
	\caption{Testbed implementation.}
	\label{fig_exp_hotmap}
	\vspace{-0.4cm}
\end{figure*}
\begin{figure*}
    \centering
    \begin{subfigure}{.2\textwidth}
        \centering
        \includegraphics[width=1.6in]{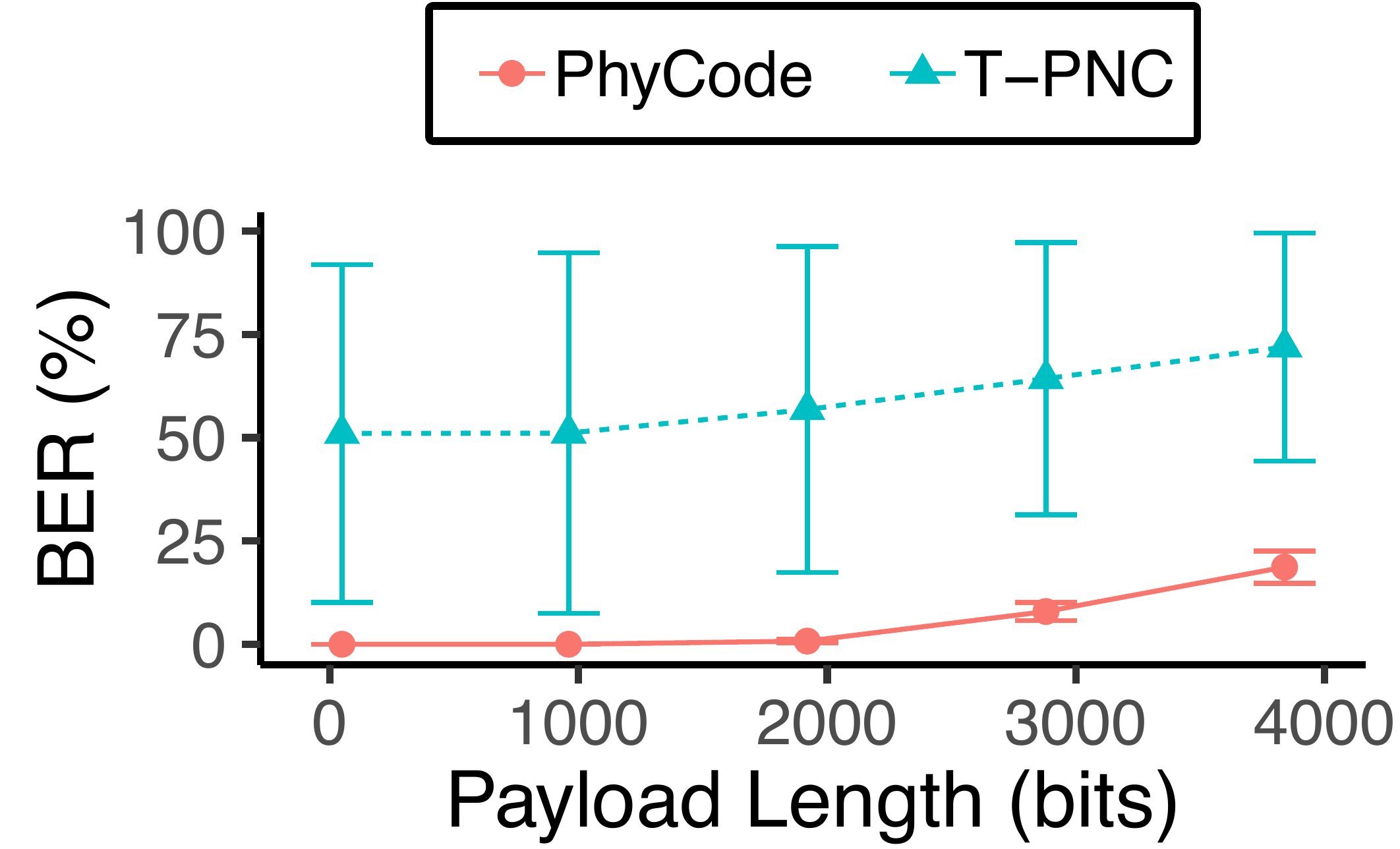}
        	\captionsetup{font={scriptsize}}
        \caption{\textbf{BER comparison}}
        \label{fig_exp_ber}
    \end{subfigure}
    \hspace{.2cm}
    \begin{subfigure}{.2\textwidth}
        \centering
        \includegraphics[width=1.6in]{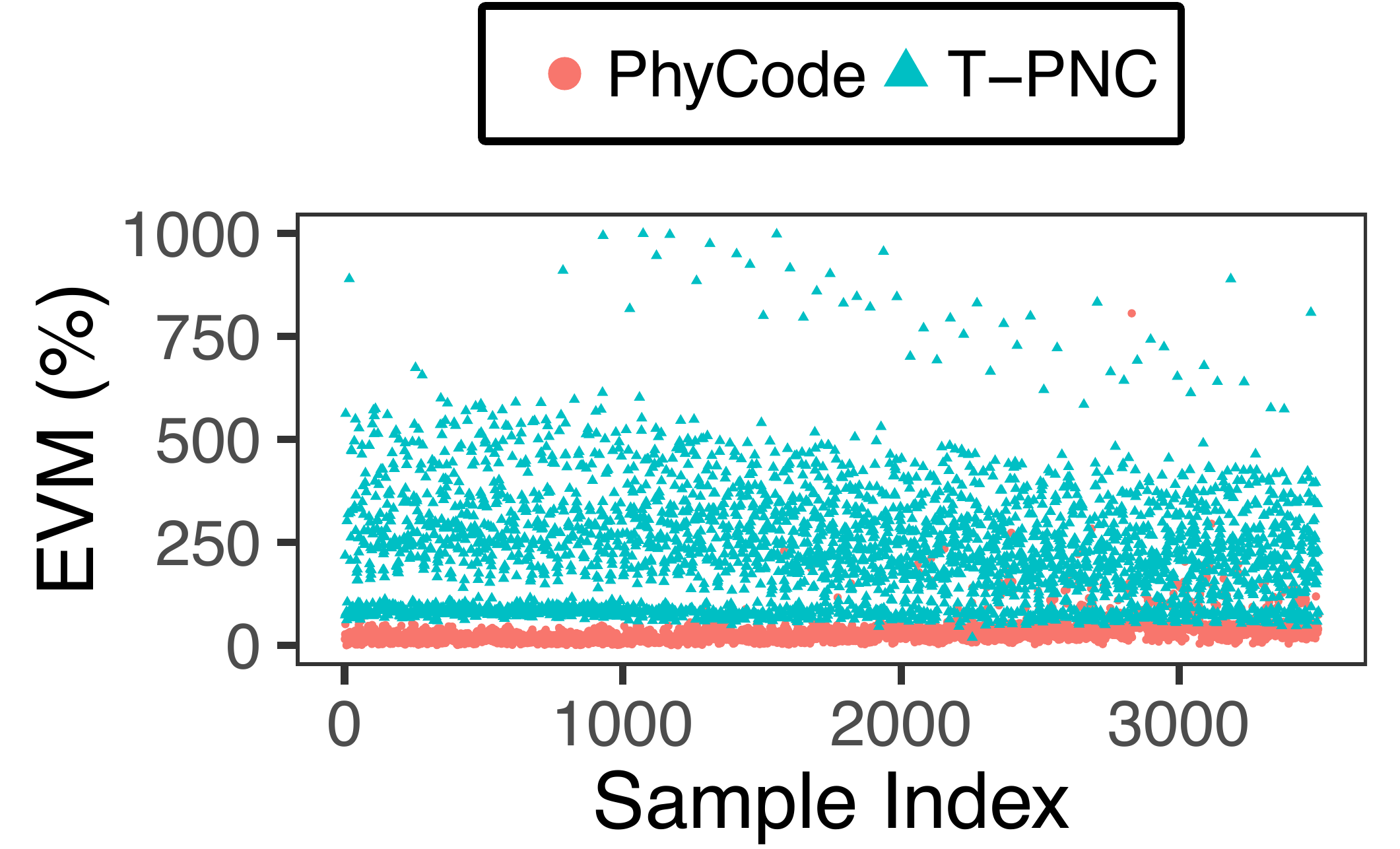}
        	\captionsetup{font={scriptsize}}
        \caption{\textbf{EVM comparison}}
        \label{fig_exp_evm}
    \end{subfigure}
    \hspace{.2cm}
    \begin{subfigure}{.2\textwidth}
        \centering
        \includegraphics[width=1.6in]{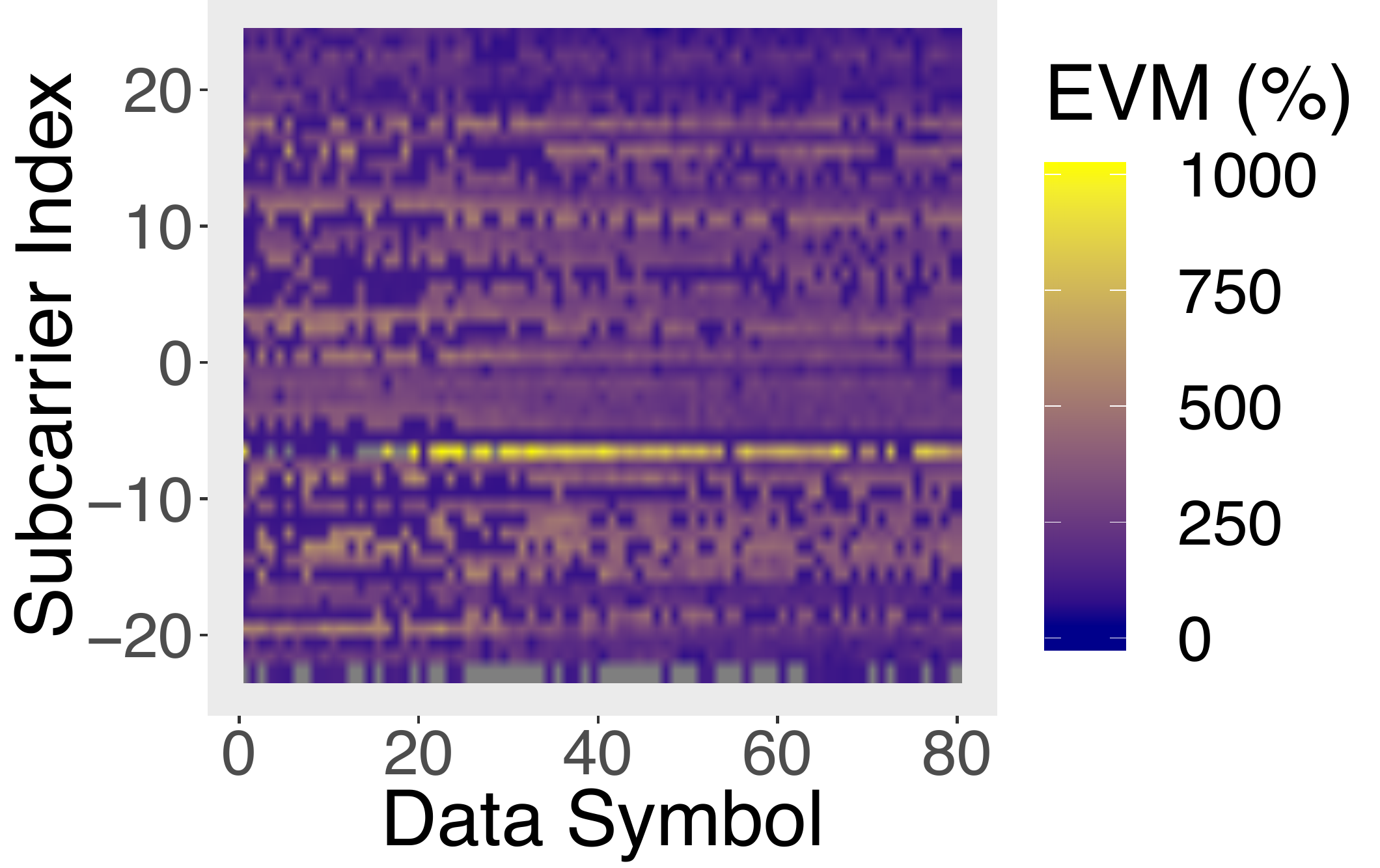}
        	\captionsetup{font={scriptsize}}
        \caption{\textbf{EVM of T-PNC (subcarriers)}}
        \label{EVM_hot_nocomp}
    \end{subfigure}
    \hspace{.2cm}
    \begin{subfigure}{.2\textwidth}
        \centering
        \includegraphics[width=1.6in]{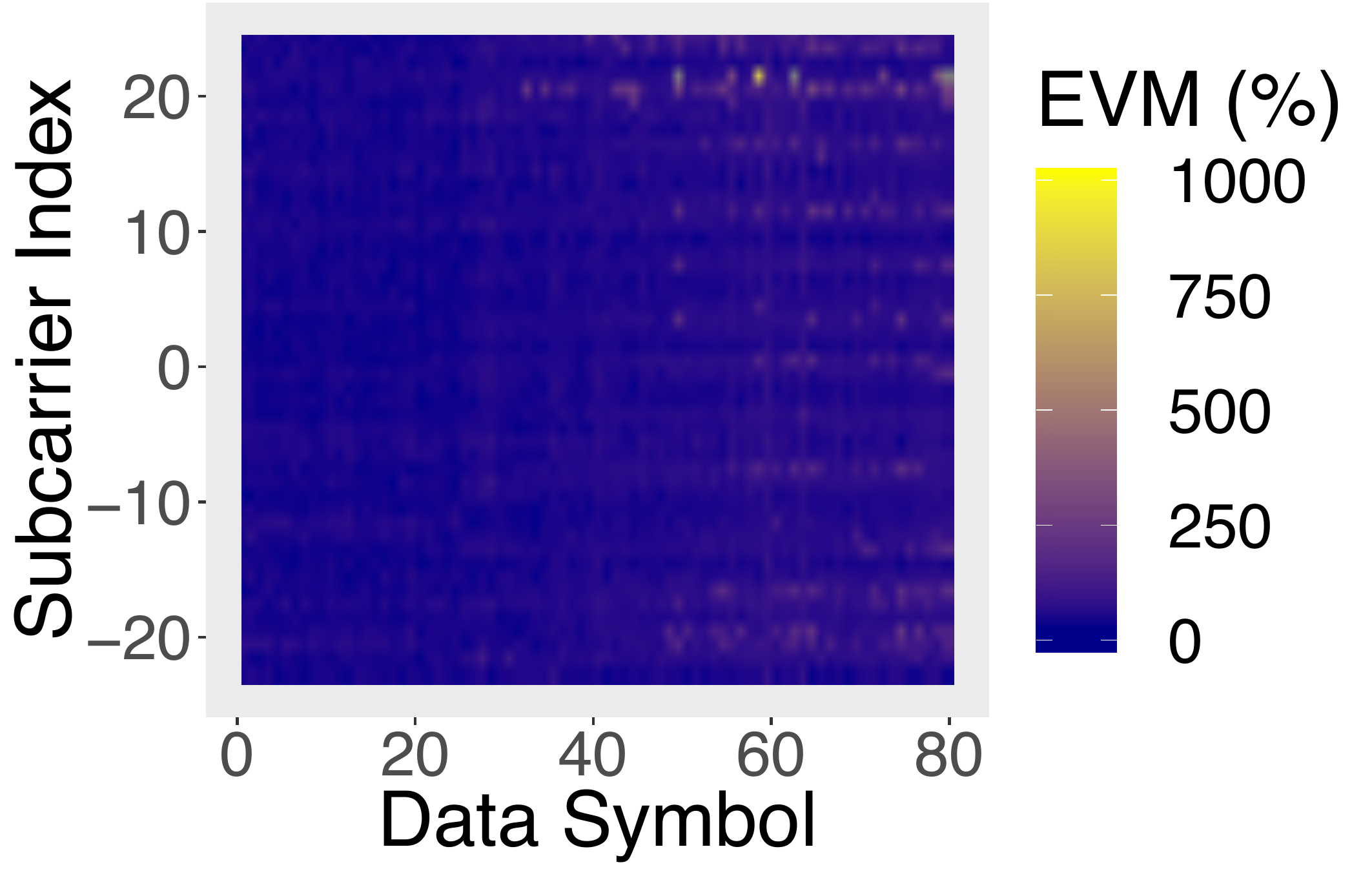}
        	\captionsetup{font={scriptsize}}
        \caption{\textbf{EVM of \titlename\ (subcarriers)}}
        \label{EVM_hot_comp}
    \end{subfigure}
	\captionsetup{font={footnotesize}}
    \caption{Performance comparison.}
     \vspace{-0.4cm}
\end{figure*}
To understand STO and SFO clearly, we emulate signals to show how the time latency affects the phase shift. Specifically, we artificially add an integer multiple of sample interval as latency to emulate STO. For SFO, we add a fractional multiple of sample interval as latency. As shown in Fig.~\ref{fig_STO_SFO}, there is a linear relationship between the subcarrier index and the phase shift. More importantly, the slope of STO is much bigger than SFO, which indicates that we can first calibrate STO and then based on this calibration result, we can zoom in to detect SFO.


\noindent \textbf{STO Calibration:} Since STO is relatively stable, we can measure and compensate it beforehand at the transmitter, just like the calibration of CFO.

\noindent \textbf{SFO Measurement:} Recall that the oscillator difference is the main cause of CFO and SFO. Hence, we can use the CFO error ratio $\epsilon= \Delta f / f$ to infer the SFO error ratio $\gamma$. Particularly, we have $\gamma \approx \epsilon$. According to Eq.~(\ref{eq_single_sfo}), we can obtain $\theta^{SFO}$ with considerable high accuracy due to the stability of CFO. However, the residual CFO remains a problem to both the CFO and SFO correction. To solve this problem, we assign two pilot samples for each source to keep tracking the CFO ($\Delta f_{\rm{step2}}$) in every symbol.

\subsection{Dynamic Decoding Scheme}
\label{sec:dyn}


So far we have corrected CFO, calibrated STO and measured SFO. The remaining part is to decode the superimposed signal. However, as we emphasized before, it is extremely challenging to correct all offsets from multiple sources simultaneously. To solve this problem, we propose a dynamic decoding scheme that changes its decoding criterion for every sample according to the measured SFO and the residual CFO. Specifically, \titlename\ combines channel condition, SFO and residual CFO together to define this dynamic decoding criterion. By using the BPSK modulation scheme as an example (Fig.~\ref{fig_constellation}), 
there are four decoding possibilities based on different channel condition $H$. In detail, $H_{1}+H_{2}$ and $-H_{1}-H_{2}$ represent ``11'' or ``00''. In contrast, $H_{1}-H_{2}$ and $-H_{1}+H_{2}$ represent ``10'' or ``01''. We can therefore decode a symbol by calculating the shortest Euclidean distance to these constellation points. 

After combining SFO and residual CFO, 
we denote $C$ as the decoding criterion which is a combination of the above channel condition $H$, SFO and residual CFO. Furthermore, by solving an optimization problem, we can decode the superimposed signal at the receiver. For simplicity, we take two transmitters as an example. Accordingly, the optimization problem can be written as follows, 
\begin{equation}
\underset{m,n} \min {\|(p_mC_{1}+p_nC_{2}) - \tilde{x} \|}_{2},
\label{eq_constellation}
\end{equation}
where $\tilde{x}$ is the symbol to be decoded, and $C_1$ and $C_2$ are the decoding criteria of the two transmitters, respectively. We denote the constellation points set for the $K$-QAM modulation as $P^K=\{p_1,p_2,...,p_{2^K}\}$. We have $p_m, p_n\in P^K$ and $m,n=1,2,...,2^K$. 
For example, in the BPSK modulation, $P^2=\{-1,1\}$. 
More generally, suppose that there are $N$ transmitters using the $K$-QAM modulation, hence, the above optimization problem can be extended as 
\begin{equation}
\underset{g_1, g_2, ..., g_N} \min {\|(p_{g_1}C_{1}+p_{g_2}C_{2}+,...,+p_{g_N}C_{N}) - \tilde{x} \|}_{2},
\end{equation}
where $p_{g_1}, p_{g_2}, ..., p_{g_N} \in P^K$ and $g_1,g_2,...,g_N=1,2,...,2^K$. 

%% file: Performance.tex
\section{Experimental Evaluation}


\subsection{Implementation}
{\bf Hardware-wise:} We implement \titlename\ on a software-defined radio platform. The hardware setup of \titlename\ is shown in Fig.~\ref{fig_exp_tx} and Fig.~\ref{fig_exp_rx}. Specifically, we use 7 Universal Software Radio Peripheral (USRP) embedded with XCVR2450 daughterboards, including three N210s and four N200s, and two USRPs connect to a PC through a Gigabit Ethernet switch. Without loss of generality, \titlename\ follows IEEE 802.11p standard, i.e. the 5.8GHz carrier frequency and 10MHz bandwidth, which can be easily applied to other OFDM related protocols. For the time synchronization, we use NI CDA-2990 as a central clock and each USRP that acts as a transmitter is connected to this clock via SMA cables.

{\bf Software-wise:} The software of \titlename\ is based on a recent Wi-Fi project programmed in GNURadio~\cite{bloessl2018performance}. In detail, we develop \titlename\ transmitters by modifying the preamble and pilot samples as described in Section~\ref{sec:preamble} and Section~\ref{sec:dyn}, respectively. 
For \titlename\ receivers, we implement the superimposed signal detection and offset compensations, such as CFO, STO and SFO, as presented in Section~\ref{sec:detection} to Section~\ref{sec:dyn}.


\subsection{Methodology}
Our goal is to evaluate the performance of \titlename\ for dealing with heterogeneous devices in dynamic environments. First, we focus on the influence of heterogeneous devices. To do this, we use power combiners and 30dB attenuators to connect two transmitters to the receiver, which can emulate a stable wireless channel in order to avoid the impact of the dynamic environment. We randomly pick up 3 out of 7 USRPs to be the transmitters and the receiver each time. This experiment was repeated 3 times in an ordinary office. Specifically, during each time, we vary the payload length from 48 bits to 4000 bits. For a fixed payload length, \titlename\ transmits packets every 5 seconds and it lasts for one hour. 
Second, we evaluate our design in dynamic environments with NLOS scenarios. To this end, we deploy our devices at 8 different locations (Fig.~\ref{fig_exp_floor}) in our building, including both LOS and NLOS scenarios. Specifically, each time we randomly choose 2  out of 8 locations (i.e., one location for the two transmitters and other location for the receiver) to deploy \titlename, and the minimum distance between the two transmitters are at least 50cm in order to form independent channels. Both transmitters send 2000 bits payload every 5 seconds.  This experiment was repeated 10 times and the total experiment lasts for 5 hours. During the experiments, people in the building just work as usual, i.e., they can either sit in their offices or walk around the corridors, which contributed to a dynamic environment. 

\subsection{Metrics}

We use the following two metrics for the evaluation: (a) Bit Error Rate (BER): the percentage of bits in error in a \titlename\ packet; (b) Error Vector Magnitude (EVM): a measure of how far the constellation points are from the ideal locations, which is a fine-grained error analysis of each sample. Both metrics are related to the decoding rate. We compare \titlename\ with the existing state-of-the-art PNC implementation~\cite{lu2013implementation,you2017reliable}, where only an average CFO was compensated to the superimposed signal. For simplicity, we denote these kinds of PNC implementation as T-PNC in the following comparison.



\subsection{Impact on Heterogeneous Devices}

We plot the comparison result of BER in Fig.~\ref{fig_exp_ber}. Obviously, \titlename\ outperforms T-PNC substantially. The underlying reason is that T-PNC suffers deeply from the offsets,  especially for the diverse behaviors of different oscillators. In contrast, \titlename\ reacts to the multiple offsets effectively. To see it more clearly, we randomly pick up one payload with 4000 bits to evaluate the result from signal constellations in a fine-grained manner. As shown in Fig.~\ref{fig_exp_evm}, the EVM result of T-PNC is scattered around and is larger than that of \titlename, which indicates the presence of a large amount of CFO. Although they compensated the signal with an average offset, the superimposed signal is still severely affected by the offsets. Furthermore, from the view of each subcarrier, we investigate the effects of STO and SFO. As shown in Fig.~\ref{EVM_hot_comp}, \titlename\ can mitigate STO and SFO effectively. However, T-PNC gets hurt from the offsets in every subcarrier as revealed in Fig.~\ref{EVM_hot_nocomp}. 


We note that \titlename\ compensates the signal well in most of the cases. But with the symbol index increasing, the damage of residual offsets becomes more obvious. In dealing with this situation, we can insert channel estimation pilot symbols periodically to ensure an accurate estimation, so we can always keep our decoding success rate at an acceptable level.

\subsection{Impact on Dynamic Environment with NLOS}
Fig.~\ref{fig_exp_scenario} shows the BER results of \titlename\ in both the LOS and NLOS scenarios, respectively. All results indicate that our design is feasible to be implemented in a practical dynamic environment with a considerable low raw BER. 

%% file: Conclusion.tex
\section{Conclusion and Future Work}
The paper introduces \titlename, a practical wireless prototype of superimposed signals in the presence of heterogeneous devices and dynamic environments, such as time synchronization errors and oscillator offsets. We demonstrate the feasibility of our design through the implementation on a software-defined radio platform. As for future work, \titlename\ can be extended to support a large number of sources concurrently transmitting signals, as a sign of the scalability, where a scheduling algorithm for the pilot sample assignment should be considered to reduce the offset tracking overhead. We can further apply the channel and error coding to improve the BER. Also, we can design a distributed time management scheme to fully support multi-source and multi-hop scenarios.

%% file: main.bbl
\begin{thebibliography}{10}
\providecommand{\url}[1]{#1}
\csname url@samestyle\endcsname
\providecommand{\newblock}{\relax}
\providecommand{\bibinfo}[2]{#2}
\providecommand{\BIBentrySTDinterwordspacing}{\spaceskip=0pt\relax}
\providecommand{\BIBentryALTinterwordstretchfactor}{4}
\providecommand{\BIBentryALTinterwordspacing}{\spaceskip=\fontdimen2\font plus
\BIBentryALTinterwordstretchfactor\fontdimen3\font minus
  \fontdimen4\font\relax}
\providecommand{\BIBforeignlanguage}[2]{{%
\expandafter\ifx\csname l@#1\endcsname\relax
\typeout{** WARNING: IEEEtran.bst: No hyphenation pattern has been}%
\typeout{** loaded for the language `#1'. Using the pattern for}%
\typeout{** the default language instead.}%
\else
\language=\csname l@#1\endcsname
\fi
#2}}
\providecommand{\BIBdecl}{\relax}
\BIBdecl

\bibitem{zhang2006hot}
S.~Zhang, S.~C. Liew, and P.~P. Lam, ``Hot topic: Physical-layer network
  coding,'' in \emph{ACM MOBICOM}, 2006.

\bibitem{katti2007embracing}
S.~Katti, S.~Gollakota, and D.~Katabi, ``Embracing wireless interference:
  Analog network coding,'' in \emph{ACM SIGCOMM}, 2007.

\bibitem{chen2017bipass}
L.~Chen, F.~Wu, J.~Xu, K.~Srinivasan, and N.~Shroff, ``Bi{P}ass: Enabling
  end-to-end full duplex,'' in \emph{ACM MOBICOM}, 2017.

\bibitem{zhang2017bi}
H.~Zhang and L.~Cai, ``Bi-directional multi-hop wireless pipeline using
  physical-layer network coding,'' \emph{IEEE Transactions on Wireless
  Communications}, 2017.

\bibitem{zhang2017design}
H.~Zhang, L.~Zheng, and L.~Cai, ``Design and analysis of hierarchical physical
  layer network coding,'' \emph{IEEE Transactions on Wireless Communications},
  2017.

\bibitem{zhang2018design}
H.~Zhang and L.~Cai, ``Design of channel coded heterogeneous modulation
  physical layer network coding,'' \emph{IEEE Transactions on Vehicular
  Technology}, 2018.

\bibitem{shannon1961two}
C.~E. Shannon, ``Two-way communication channels,'' in \emph{4th Berkeley
  Symposium on Math. Stat. and Prob.}, 1961.

\bibitem{lu2013implementation}
L.~Lu, T.~Wang, S.~C. Liew, and S.~Zhang, ``Implementation of physical-layer
  network coding,'' \emph{Physical Communication}, 2013.

\bibitem{kong2015mzig}
L.~Kong and X.~Liu, ``m{Z}ig: Enabling multi-packet reception in {Z}ig{B}ee,''
  in \emph{ACM MOBICOM}, 2015.

\bibitem{saito2013non}
Y.~Saito, Y.~Kishiyama, A.~Benjebbour, T.~Nakamura, A.~Li, and K.~Higuchi,
  ``Non-orthogonal multiple access ({NOMA}) for cellular future radio access,''
  in \emph{IEEE VTC}, 2013.

\bibitem{you2017reliable}
L.~You, S.~C. Liew, and L.~Lu, ``Reliable physical-layer network coding
  supporting real applications,'' \emph{IEEE Transactions on Mobile Computing},
  2017.

\bibitem{meyr1997digital}
H.~Meyr, M.~Moeneclaey, and S.~Fechtel, \emph{Digital Communication Receivers:
  Synchronization, Channel Estimation, and Signal Processing}.\hskip 1em plus
  0.5em minus 0.4em\relax John Wiley \& Sons, Inc., 1997.

\bibitem{lombardi2002fundamentals}
M.~A. Lombardi, ``Fundamentals of time and frequency,'' \emph{The Mechatronics
  Handbook}, 2002.

\bibitem{mahinthan2008partner}
V.~Mahinthan, L.~Cai, J.~W. Mark, and X.~Shen, ``Partner selection based on
  optimal power allocation in cooperative-diversity systems,'' \emph{IEEE
  Transactions on Vehicular Technology}, 2008.

\bibitem{you2015network}
L.~You, S.~C. Liew, and L.~Lu, ``Network-coded multiple access {II}: Toward
  real-time operation with improved performance,'' \emph{IEEE Journal on
  Selected Areas in Communications}, 2015.

\bibitem{hamed2018chorus}
E.~Hamed, H.~Rahul, and B.~Partov, ``Chorus: Truly distributed
  distributed-{MIMO},'' in \emph{ACM SIGCOMM}, 2018.

\bibitem{rahul2010sourcesync}
H.~Rahul, H.~Hassanieh, and D.~Katabi, ``Source{S}ync: A distributed wireless
  architecture for exploiting sender diversity,'' in \emph{ACM SIGCOMM}, 2010.

\bibitem{vasisht2016decimeter}
D.~Vasisht, S.~Kumar, and D.~Katabi, ``Decimeter-level localization with a
  single {WiFi} access point.'' in \emph{USENIX NSDI}, 2016.

\bibitem{wang2012efficient}
J.~Wang, H.~Hassanieh, D.~Katabi, and P.~Indyk, ``Efficient and reliable
  low-power backscatter networks,'' in \emph{ACM SIGCOMM}, 2012.

\bibitem{jin2017fliptracer}
M.~Jin, Y.~He, X.~Meng, Y.~Zheng, D.~Fang, and X.~Chen, ``Fliptracer: Practical
  parallel decoding for backscatter communication,'' in \emph{ACM MOBICOM},
  2017.

\bibitem{jin2018parallel}
M.~Jin, Y.~He, X.~Meng, D.~Fang, and X.~Chen, ``Parallel backscatter in the
  wild: When burstiness and randomness play with you,'' in \emph{ACM MOBICOM},
  2018.

\bibitem{eletreby2017empowering}
R.~Eletreby, D.~Zhang, S.~Kumar, and O.~Ya{\u{g}}an, ``Empowering low-power
  wide area networks in urban settings,'' in \emph{ACM SIGCOMM}, 2017.

\bibitem{hessar2018netscatter}
M.~Hessar, A.~Najafi, and S.~Gollakota, ``Netscatter: Enabling large-scale
  backscatter networks,'' in \emph{USENIX NSDI}, 2019.

\bibitem{tse2005fundamentals}
D.~Tse and P.~Viswanath, \emph{Fundamentals of Wireless Communication}.\hskip
  1em plus 0.5em minus 0.4em\relax Cambridge {U}niversity {P}ress, 2005.

\bibitem{schmitz2017distributed}
J.~Schmitz, F.~Bartsch, M.~Hern{\'a}ndez, and R.~Mathar, ``Distributed software
  defined radio testbed for real-time emitter localization and tracking,'' in
  \emph{IEEE Communications Workshops (ICC Workshops)}, 2017.

\bibitem{GollakotaZigzag}
S.~Gollakota and D.~Katabi, ``Zig{Z}ag decoding: Combating hidden terminals in
  wireless networks,'' in \emph{ACM SIGCOMM}, 2008.

\bibitem{bloessl2018performance}
B.~Bloessl, M.~Segata, C.~Sommer, and F.~Dressler, ``Performance assessment of
  {IEEE} 802.11p with an open source {SDR}-based prototype,'' \emph{IEEE
  Transactions on Mobile Computing}, 2018.

\bibitem{schmidl1997robust}
T.~M. Schmidl and D.~C. Cox, ``Robust frequency and timing synchronization for
  {OFDM},'' \emph{IEEE Transactions on Communications}, 1997.

\end{thebibliography}
